\documentclass[reprint,onecolumn,superscriptaddress,nofootinbib,amsmath,amssymb, aps,prl]{revtex4}

\usepackage[a4paper, margin=2.50cm]{geometry}

\usepackage{bbm}
\usepackage{graphicx}
\usepackage{fancyhdr, color}
\usepackage{hyperref} 

\def\N{{\rm N}}

\def\la{\langle}
\def\ra{\rangle}
\def\om{\omega}
\def\Om{\Omega}

\def\tr{{\rm tr}}

\newcommand{\beq}{\begin{equation}}
\newcommand{\eeq}{\end{equation}}
\newcommand{\beqa}{\begin{eqnarray}}
\newcommand{\eeqa}{\end{eqnarray}}

 \newcommand{\chqre}{{\color[rgb]{0.4,0.2,0.9} \sc Qbook:Ch.43}} 

\begin{document}

\fancyhead[C]{\sc \color[rgb]{0.4,0.2,0.9}{Quantum Thermodynamics book}}
\fancyhead[R]{}
\title{Friction-free quantum machines}


\author{Adolfo del Campo}
\email{adolfo.delcampo@umb.edu} 
\affiliation{Department of Physics, University of Massachusetts, Boston, MA 02125, USA}
\affiliation{Theory Division, Los Alamos National Laboratory, MS-B213, Los Alamos, NM 87545, USA}
\author{Aur\'elia Chenu}
\affiliation{Theory Division, Los Alamos National Laboratory, MS-B213, Los Alamos, NM 87545, USA}
\author{Shujin Deng}
\affiliation{State Key Laboratory of Precision Spectroscopy, East China Normal University, Shanghai 200062, P. R. China}
\author{Haibin Wu}
\affiliation{State Key Laboratory of Precision Spectroscopy, East China Normal University, Shanghai 200062, P. R. China}

\date{\today}

\begin{abstract}
The operation of a quantum heat engine in finite time generally faces a trade-off between efficiency and power. 
Using shortcuts to adiabaticity (STA), this trade off can be avoided to engineer thermal machines that operate at maximum efficiency and tunable output power. We demonstrate the use of STA to engineer a scalable superadiabatic quantum Otto cycle and report recent experimental progress to tailor quantum friction in finite-time quantum thermodynamics. In the presence of quantum friction, it is also shown that the use of a many-particle working medium can boost the performance of the quantum machines with respect to an ensemble of single-particle thermal machines.

\end{abstract}

\maketitle

\thispagestyle{fancy}


\section{Heat engines today}
Catalyzing the industrial revolution, heat engines have played a decisive role in the history of humankind. 
Their study also led to a paradigm shift that transformed physics, setting the ground for a new type of science beyond Newtonian mechanics in the nineteenth century. Together with the study of heat conduction by Fourier, heat engines led to  the birth of thermodynamics. 
The quantification of the efficiency of an engine driven through a thermodynamic cycle  led to the notion of irreversibility, which has been described as the most original contribution of this field \cite{Prigogine84}.

Over the last two centuries, technological advances have motivated the thermodynamic description of  increasingly smaller systems, including chemical processes, single biomolecules and simple quantum systems. At the nanoscale, thermal and quantum fluctuations appear as a dominant new ingredient.

 In parallel with these developments, a number of fundamental problems, ranging from blackhole physics to the cost of computation, have unravelled the role of information. Quantum thermodynamics flourishes currently in this context merging notions of foundations of physics, information theory and statistical mechanics  \cite{bookQThermo04,VA15,Goold16}.

In the quantum domain, the study of  heat engines keeps facilitating further advances, building on  the dialogue between fundamental questions and applied science. 
A full quantum description of heat engines was first put forward in \cite{Alicki79,Kosloff84}. 
It is now understood that information can be used as a resource from which to extract work, leading to the notion of information-driven heat engines such as that proposed by Szilard \cite{LR03,Kim11}. 
Quantum optical devices including masers and lasers \cite{Scovil59},  together with physical processes such as  light harvesting in natural and artificial systems \cite{Dorfman13,Scully10,Scully11,Creatore13}, can be analyzed in terms of quantum thermodynamic cycles. One may therefore hope for an interplay between quantum thermodynamics and energy science.

A central result in thermodynamics is that the efficiency of any  heat engine  run using  two equilibrium thermal reservoirs with inverse temperatures $\beta_c$ and $\beta_h$ (satisfying $\beta_c>\beta_h$) has as a universal upper bound the Carnot efficiency $\eta_{C}=1-\beta_h/\beta_c$. This bound however may be reached only in engines that run infinitely slowly.
From the macroscale to quantum world, realistic heat engines are expected to operate in finite time. 
This necessity comes from an observed trade-off between efficiency and power, that has sometimes been referred to as ``the tragedy of finite time thermodynamics". While the maximum efficiency of a thermodynamic cycle, $\eta_{\rm max}\leq\eta_C$, can be reached in principle under sufficiently slow driving, the practical desireratum of a finite output power sets an upper bound to the operation time.  As a result, an attempt to increase the output power by reducing the cycle operation time leads to the emergence of friction and the reduction of the cycle efficiency. The optimization of this tradeoff has motivated a substantial body of literature. Pioneering works on the finite-time thermodynamics  of classical heat engines established that the performance at maximum power is characterized  by the so-called Curzon-Ahlborn efficiency  
$\eta_{CA}=1-\sqrt{1-\eta_C}$ \cite{CA75}. Finite-time  thermodynamics was consolidated as an important subject area of research, see, e.g., \cite{FTTbook1,Salamon01}.  In the quantum domain, following the work by Scovil and Schulz-DuBois \cite{Scovil59},  early descriptions of quantum heat engines emphasized their operation in finite time. Subsequent works accumulated evidence indicating that the trade off between efficiency and power holds as well in the quantum realm \cite{GK92,KF02,FK03,YK06,Salamon09,Rezek09,Abah12,AL16}. 

Over the last few years, however, it has been shown that this tradeoff is not fundamental and can be avoided. Even more, it is possible in principle to operate a heat engine at maximum efficiency and  high output power \cite{Deng13,delcampo14,Beau16,Chotorlishvili16,AL17,Deng18,Li18}. Such a frictionless quantum heat engine may be engineered using shortcuts to adiabaticity (STA):  nonadiabatic protocols that lead to the same final state that would be achieved under slow adiabatic driving \cite{STAR}. In addition, nonadiabatic many-particle effects have been shown to  boost the performance of quantum thermal machines \cite{Jaramillo16}. This chapter summarizes these developments, focusing on the description of quantum heat engines (QHEs) with an emphasis on their operation in finite time.

\section{Trapped quantum fluids as working media}

The substance that performs work and on which work is also done in different stages of a thermodynamic cycle is generally referred to as the working substance or working medium. In classical thermodynamics, the  performance of a heat engine is largely independent of its choice. Theoretical models of quantum heat engines have shown that this conclusion often holds in the quantum domain under slow driving, but not generally \cite{ZP15,Bengtsson18}. In addition, the operation of the engine away from equilibrium generally exhibits specific signatures of the working medium.  This may result from using nonthermal reservoirs or operating the cycle in finite-time, i.e., in a non-adiabatic fashion. The quest for quantum effects that boost  the performance in heat engines arises in this scenario as a natural pursuit. The choice of the working substance is further guided by the range of available quantum platforms, simplicity and aesthetic appeal. Quantum systems with discrete energy levels or a continuous spectra can be considered.

In what follows, we shall focus on the realization of heat engines with confined particles in a time-dependent harmonic trap as a working medium \cite{Jaramillo16}. 
In particular, we consider the  family of quantum many-body systems with Hamiltonian
 \begin{equation}
\hat{H}(t) =\sum_{i=1}^{\N}\left[-\frac{\hbar^2}{2m}\nabla_i^2+\frac{1}{2}m\omega(t)^2 {\bf r}_{i}^{2}\right]+
\sum_{i<j}V({\bf r}_i-{\bf r}_j)\ ,
 \label{Hsu11}
\end{equation}
 that describes  $\N$  particles of equal mass $m$ in an isotropic harmonic trap with frequency  $\omega(t)$.
It will prove convenient to choose the pairwise interaction $V({\bf r})$ as a homogeneous function of degree $-2$, satisfying,
 \beqa
 V(b{\bf r})=b^{-2}V({\bf r})\ . 
 \eeqa
This choice accounts for a  variety of many-body systems including free noninteracting gases,  quantum fluids with hard-core interactions, inverse-square interactions models, as well as Bose gases with s-wave contact interactions in two spatial dimensions ($d=2$),   among other  examples.
It further simplifies the dynamics under a modulation of the trapping frequency $\om(t)$, due to the emergence of scale-invariance as a dynamical symmetry \cite{Gambardella75,delcampo11b,
DB12,delcampo13,DJD14}.

In the adiabatic limit achieved under slow driving of $\om(t)$, scale invariance determines the instantaneous mean energy 
\beqa
\la \hat{H}(t)\ra_{\rm ad}=\frac{1}{b_{\rm ad}^2}\la \hat{H}(0)\ra\ ,
\eeqa
where the adiabatic scaling factor is given by
\beqa
b_{\rm ad}=\sqrt{\om(0)/\om(t)} \ .
\eeqa

Under arbitrary driving, the nonadiabatic mean energy is fixed by the initial mean energy, particle position fluctuations and squeezing according to \cite{Jaramillo16}
\beqa
\la \hat{H}(t)\ra=\frac{1}{b^2}\la \hat{H}(0)\ra-\frac{m}{2}\left(b\ddot{b}
-\dot{b}^2\right)\sum_{i=1}^\N \la {\bf r}_i^2(0)\ra+\frac{\dot{b}}{2b}\sum_{i=1}^\N\la \{ {\bf r}_i,{\bf p}_i \}(0)\ra\ .
\eeqa
Here, the time-dependent coefficients are fixed by the scaling factor, that is a solution of the Ermakov equation
\beqa
\label{EPE}
\ddot{b}+\om(t)^2b=\om_0^2b^{-3}\ .
\eeqa
Provided that the initial state is at thermal equilibrium, so that the squeezing term vanishes, the nonadiabatic mean-energy following a variation of the trapping frequency $\om(t)$ is given by the relation
\beqa
\label{aven0}
\la \hat{H}(t)\ra =Q^{\ast}(t) \la \hat{H}(t)\ra_{\rm ad}\ ,
\eeqa
where the nonadiabatic factor $Q^{\ast}\geq 1$ reads
\beqa
\label{Qstar}
Q^{\ast}(t)&=&\frac{\om_0}{\om(t)}\left(\frac{1}{2b^2}+\frac{\om(t)^2}{2\om_0^2}b^2+\frac{\dot{b}^2}{2\om_0^2}\right)\ .
\eeqa
Here $b(t)$ is the solution of (\ref{EPE}) subject to the boundary conditions $b(0)=1$ and $\dot{b}(0)=0$, to account for the initial equilibrium state. 
The nonadiabatic factor $Q^{\ast}$ was first discussed by Husimi in the study of a single-particle driven oscillator \cite{Husimi53}. As we shall see, 
values of $Q^{\ast}>1$  limit the finite-time efficiency of thermodynamic cycles.
We note that in the adiabatic limit time derivatives of the scaling factor can be ignored and 
\beqa
b(t)&\rightarrow& b_{\rm ad}=\sqrt{\om_0/\om(t)}\ ,\\
Q^{\ast}(t)&\rightarrow& Q_{\rm ad}^{\ast}(t)=1\ .
\eeqa

In the next section, we shall describe a many-particle QHE  based on an Otto cycle and consider thermal equilibrium states in the canonical ensemble at (inverse) temperature $\beta$ described by a density matrix of the form
\beqa
\hat{\rho}=\frac{e^{-\beta  \hat{H}}}{\tr(e^{-\beta  \hat{H}})}\ .
\eeqa

\section{The quantum Otto cycle: finite-time thermodynamics}

\begin{figure}
\begin{center}
\includegraphics[width=0.6\linewidth]{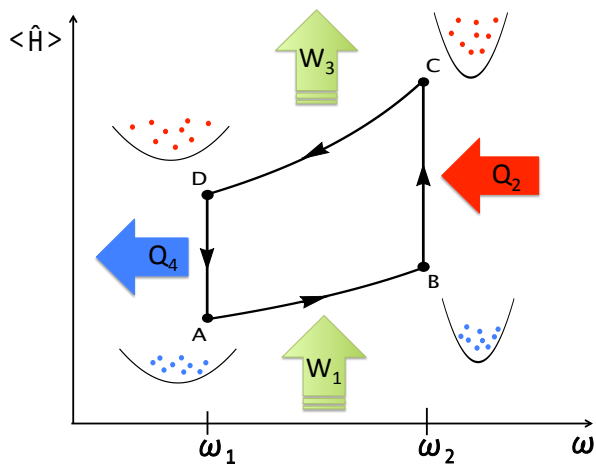}
\end{center}
\caption{\textbf{Many-particle quantum heat engine.} A quantum Otto cycle  realized by a quantum fluid as a working medium involves expansion and compression processes in which  the frequency of the harmonic trap varies  between $\om_1$ and $\om_2$.  These strokes  in which the dynamics is unitary are alternated with heating and cooling processes in which the system Hamiltonian is kept constant and heat is either absorbed or released. }
\label{figure1}
\end{figure}
The quantum Otto cycle consists of four  consecutive strokes performed on  a quantum system that acts as a working substance \cite{KR17}, as shown in Figure \ref{figure1}:

1) In a first isentropic  stroke ($A\rightarrow B$), the working substance  prepared in an equilibrium state  at low temperature 
undergoes a compression induced by increasing the trap frequency $\omega(t)$. In this stroke the dynamics is unitary as there is no coupling to the environment. Thus, there is no heat exchange and entropy is conserved. The change in the internal energy constitutes the work input. Upon completion of the stroke, the final state  is generally nonthermal and out of equilibrium.

2) Subsequently, the system is heated up at constant volume to a new thermal state at high temperature ($B\rightarrow C$). 
As the Hamiltonian of the working substance is kept constant, the change of internal energy is exclusively given by the heat absorbed, and no work is done in this stroke. Once at equilibrium the system is decoupled from the hot reservoir. This assumes that no work is done by coupling and decoupling the working substance to and from the hot reservoir, an approximation which can be called into question beyond the weak coupling limit.

3) The equilibrium state at inverse temperature $\beta_h$ is taken as the starting point of an expansion stroke ($C\rightarrow D$) in which the dynamics is unitary and the work output is set by the change in the internal energy of the working substance, that is found in a nonequilibrium state upon completion of the stroke.

4) The cycle is then completed by a second isochoric stroke ($D\rightarrow A$) in which the working substance thermalized to the low-temperature equilibrium state at $\beta_c$. Again, in this stroke it  is assumed that the work done by putting in contact the working substance with the cold reservoir is negligible. The Hamiltonian of the working substance can then be considered constant and the heat released is therefore given by the change in the energy of the working substance.

The efficiency of a heat engine is defined as the total output work per input heat
\begin{eqnarray}\label{Efficiency}
\eta=-\frac{\langle W_1\rangle+\langle W_3\rangle}{\langle {\rm Q}_2\rangle}\ ,
\end{eqnarray}
where $\langle W_{1(3)}\rangle=\langle \hat{H}\rangle_{B(D)}-\langle \hat{H}\rangle_{A(C)}$ and $\langle {\rm Q}_{2(4)}\rangle=\langle \hat{H}\rangle_{C(A)}-\langle \hat{H}\rangle_{B(D)}$.  As the working substance is decoupled from the thermal reservoir during the isentropic strokes, when the dynamics is unitary, we use the scaling dynamics \eqref{Hsu11} to predict the mean energy according to \eqref{aven0}. Direct computation yields
\begin{subequations}\label{WQWQ}
\begin{gather}
 \langle W_1\rangle = \langle \hat{H}\rangle_B-\langle \hat{H}\rangle_A = (Q_{AB}^*\frac{\omega_2}{\omega_1}-1)\langle \hat{H}\rangle_A\ ,\label{WQ1} \\
 \langle W_3\rangle = \langle \hat{H}\rangle_D-\langle \hat{H}\rangle_C = (Q_{CD}^*\frac{\omega_1}{\omega_2}-1)\langle \hat{H}\rangle_C\ , \label{WQ2}\\
 \langle \mathrm{Q}_2\rangle= \langle \hat{H}\rangle_C-\langle \hat{H}\rangle_B = \langle \hat{H}\rangle_C - Q_{AB}^*\frac{\omega_2}{\omega_1}\langle \hat{H}\rangle_A\ ,\label{WQ3}\\
 \langle \mathrm{Q}_4\rangle= \langle \hat{H}\rangle_D-\langle \hat{H}\rangle_A = Q_{CD}^*\frac{\omega_1}{\omega_2}\langle \hat{H}\rangle_C - \langle \hat{H}\rangle_A\ ,\label{WQ4}
\end{gather}
\end{subequations}
where $\langle \hat{H}\rangle_A$ and $\langle \hat{H}\rangle_C$ denote the equilibrium energies of the thermal states parameterized by $(\omega_1,\beta_{\rm c})$ and $(\omega_2,\beta_{\rm h})$, respectively. A similar analysis  was  first reported in the study of a quantum heat engine with a single-particle harmonic oscillator as a working medium \cite{YK06,Abah12}. Equations (\ref{WQWQ}) hold however for  the family of many-particle quantum systems described by the Hamiltonian class (\ref{Hsu11})  \cite{Jaramillo16,Beau16}. Further, $Q_{AB}^*$ and $Q_{CD}^*$ denote the nonadiabatic factors at the end of the compression and expansion strokes, respectively.  Expressions (\ref{WQWQ}) assume that no work is done by coupling and decoupling the working medium to and from the heat reservoirs, e.g., in the  limit of weak coupling between the working substance and the heat reservoirs.
The efficiency of the many-particle quantum heat engine run in finite time is then given by
\begin{eqnarray}\label{Efficiency}
\eta=1-\frac{\omega_1}{\omega_2}\left(\frac{Q_{CD}^{\ast}\langle \hat{H}\rangle_C-\frac{\omega_2}{\omega_1}\langle \hat{H}\rangle_A}{\langle \hat{H}\rangle_C-Q_{AB}^{\ast}\frac{\omega_2}{\omega_1}\langle \hat{H}\rangle_A}\right)\ .
\end{eqnarray}
In the adiabatic limit ($Q^{\ast}_{AB(CD)}\rightarrow 1$)  the engine operates at the maximum Otto efficiency 
\beqa
\eta_{\rm max}=\eta_O=1-\frac{\omega_1}{\omega_2}\ ,
\eeqa
which is shared as an upper bound by both single- and many-particle quantum and classical Otto cycles. 
Another relevant limit corresponds to the sudden quench of the trapping frequency between $\om_1$ and $\om_2$. The nonadiabatic factor $Q^*_{\text{sq}}=(\omega_1^2+\omega_2^2)/(2\omega_1\omega_2)$ \cite{Husimi53} is symmetric on $\om_1$ and $\om_2$. As a result, it describes both a sudden compression and expansion strokes.

When a  compression (expansion) of finite duration $\tau$ is considered in which the frequency  varies monotonically as a function of time we have $Q^{\ast}_{AB(CD)}\leq Q^{\ast}_{\rm sq}$. 
Equation \eqref{Efficiency} then implies that the finite-time efficiency $\eta$ is bounded from below and above as \cite{Jaramillo16}
\begin{equation}\label{efficiencybound1}
\eta_{\text{sq}}\leq \eta \leq \eta_O\ ,
\end{equation}
where $\eta_{\text{sq}}$ is the efficiency under a  sudden quench.  For a monotonic frequency modulation, we have that $Q^*(\tau)\rightarrow Q^*_{\text{sq}}$ in the sudden-quench limit $\tau\rightarrow 0$.  
In addition,  as proved in \cite{Jaramillo16},  the efficiency \eqref{Efficiency} is bounded from above by a non-adiabatic Otto limit,
\begin{equation}\label{efficiencybound2}
\eta \leq 1-Q_{CD}^{\ast}\frac{\omega_1}{\omega_2}\ ,
\end{equation}
that is independent of the number of particles $\N$ and interaction potential $V$.

This formula encodes the ``tragedy of finite-time thermodynamics'' in the many-particle setting: 
The maximum efficiency is achieved under slow driving, in the adiabatic limit,
when the QHE operates at 
vanishing output power $-(\la W_1\ra+\la W_3\ra)/\tau_c$ as a result of the requirement for a long cycle time $\tau_c$. 
By contrast, realistic engines operated in finite time achieve a finite output power at the cost of introducing 
nonadiabatic energy excitations that represent  quantum friction and lower the efficiency of the cycle.  Note that for the characterization of the engine performance, energy excitations can be associated with friction even if the dynamics along the expansion and compression strokes is unitary. That nonadiabatic effects generally decrease the engine efficiency follows from the fact that $Q_{AB(CD)}^{\ast}\geq 1$ and the expression for the finite-time efficiency (\ref{Efficiency}).
However, we will show that this trade-off is not fundamental in nature and can be avoided.

\section{Shortcuts to adiabaticity: Towards a superadiabatic Otto cycle}

The efficiency of a quantum Otto cycle is limited by the presence of quantum friction generated in the nonadiabatic dynamics of the isentropic strokes. 
However, there exist nonadiabatic protocols in which the quantum friction vanishes upon completion of the stroke. To demonstrate this we consider the dynamics of an isentropic stroke in which  the working substance is described by the Hamiltonian (\ref{Hsu11}). At the beginning of the stroke, the working substance is  in equilibrium 
and $Q^*(0)=1$. We consider the duration of the stroke to be $\tau$ and look for a shortcut protocol
for which quantum friction vanishes upon completion of the stroke, this is, $Q^*(\tau)=1$. We shall refer to such a protocol as a shortcut to adiabaticity (STA) \cite{STAR}. More generally, a STA is any  fast nonadiabatic protocol that provides an alternative to adiabatic evolution, leading to the same final state without the requirement of slow driving. 

Whenever the evolution of the working substance exhibits scale invariance, it is possible to design a STA by reverse engineering the dynamics. This approach was first discussed for the single-particle harmonic oscillator in \cite{Chen10} and extended to driven quantum fluids in \cite{delcampo11b}. Our strategy to design STA, however, focuses  on the analysis of the
nonadiabatic factor $Q^*$ in Eq. (\ref{Qstar}) that plays the role of quantum friction.
One first singles out a trajectory of the scaling factor $b(t)$ connecting the initial and final states, both at equilibrium. 
From the explicit expression  for $Q^*$, Eq. (\ref{Qstar}), we identify the boundary conditions,
\begin{eqnarray}
b(0) &=& 1\  ,   \qquad \qquad b(\tau) = b_\tau\ , \\
\dot{b}(0) &=&0\  ,   \qquad \qquad \dot{b}(\tau) =0\ , \\
\ddot{b}(0) &=&0\  ,   \qquad \qquad \ddot{b}(\tau) = 0\ ,
\end{eqnarray}
where the vanishing of $\ddot{b}$ at the end points $t=0,\tau$ is optional and imposed for smoothness of the associated frequency modulation, to be determined. Here, $ b_\tau$ is the expansion or compression scaling factor upon completion of the stroke.
Making use of them it is possible to fix the form of an interpolating ansatz of the form $b(t) = \sum_{n=0}^5 c_n \left( \frac{t}{\tau} \right)^n $, i.e.,
\begin{equation}
b(t) = 1 + 10 (b_{\tau}-1) \left(\frac{t}{\tau} \right)^3 
- 15 (b_{\tau}-1) \left(\frac{t}{\tau} \right)^4
+6  (b_{\tau}-1) \left(\frac{t}{\tau} \right)^5\ .
\end{equation}
Having found a trajectory $b(t)$ of the scaling factor associated with a STA with  $Q^*(\tau)=1$, we determine the required modulation of the driving frequency from the Ermakov equation (\ref{EPE}), as
\beqa
\om^2(t)=\frac{\om_0^2}{b^4(t)}-\frac{\ddot{b}(t)}{b(t)}\ .
\eeqa
Thus, a nonadiabatic isentropic stroke with this modulation of the trapping frequency is free from friction.

There are closely related approaches to engineer STA. One strategy involves choosing a reference modulation of the trapping frequency $\om(t)$ assuming adiabatic dynamics. Setting $\ddot{b}=0$ in the Ermakov equation one finds the corresponding adiabatic scaling factor to be given by $b_{\rm ad}=[\om(0)/\om(t)]^{1/2}$. A STA by local counterdiabatic driving can then be implemented by changing  the trap with the modified frequency \cite{delcampo11epl,delcampo13,DJD14}
\beqa
\Om(t)^2=\om(t)^2-\frac{\ddot{b}_{\rm ad}}{b_{\rm ad}}=\om(t)^2-\frac{3}{4}\left[\frac{\dot{\om}(t)}{\om(t)}\right]^2+\frac{\ddot{\om}(t)}{2\om(t)}\ .
\eeqa
For completeness we also note that scaling laws are not restricted to harmonic traps and can occur in other confinements with specific space and time dependencies \cite{DJD14}. 
For instance, in a box-like trap, scaling invariance occurs provided that a time-dependent harmonic trap is superimposed during the expansion of the box trap \cite{DB12}.  
Currently, there is a variety of experimental techniques that allow for the engineering of arbitrary trapping potentials $V({\bf r},t)$ for ultracold atoms, including the use of time-averaged potentials \cite{Henderson09,Bell16} and digital micromirror devices (DMDs) \cite{Gauthier16}.

\begin{figure}[t]
	\begin{center}
	{\includegraphics[width = 0.3 \textwidth]{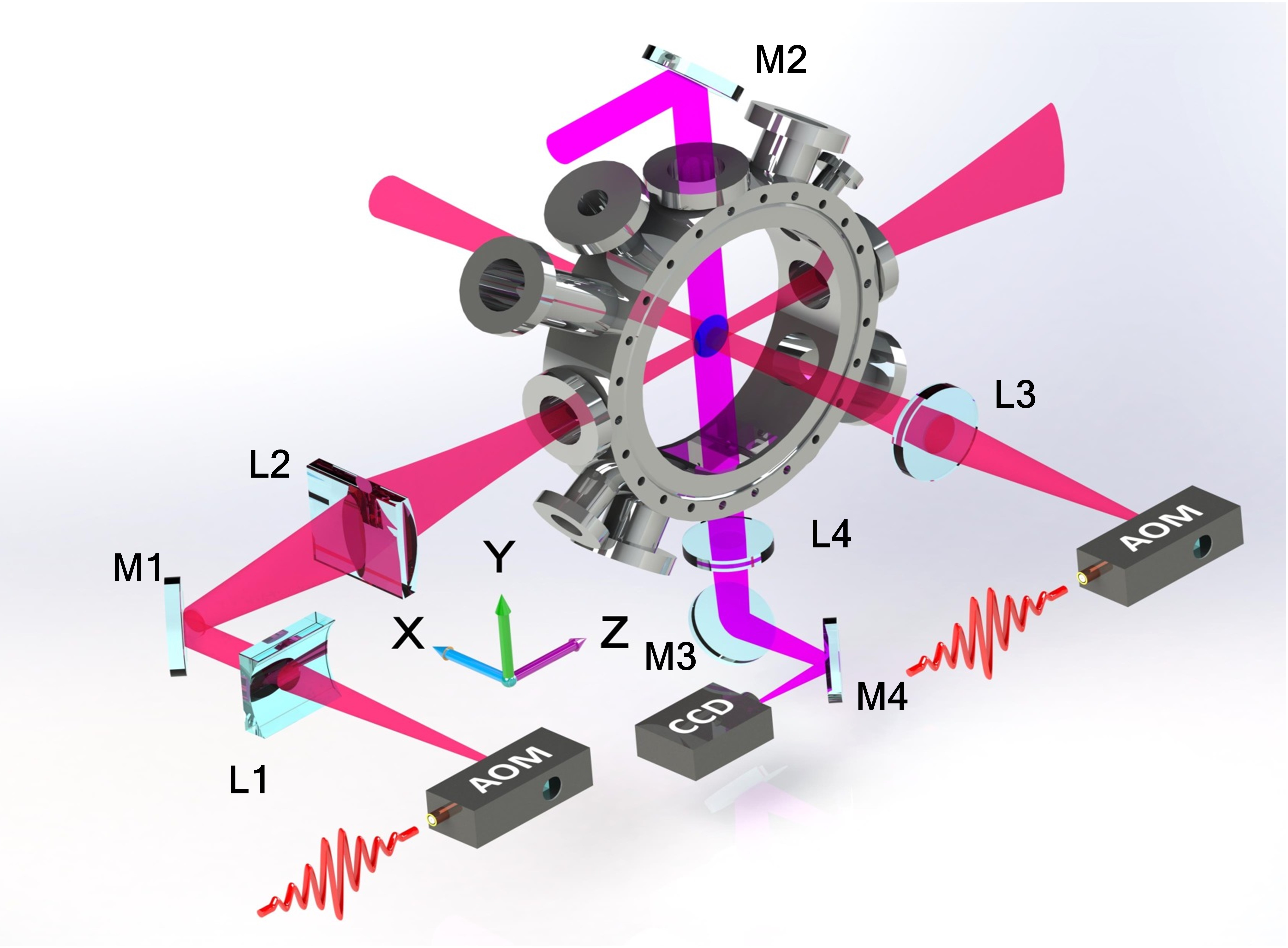}}
        {\includegraphics[width = 0.3 \textwidth]{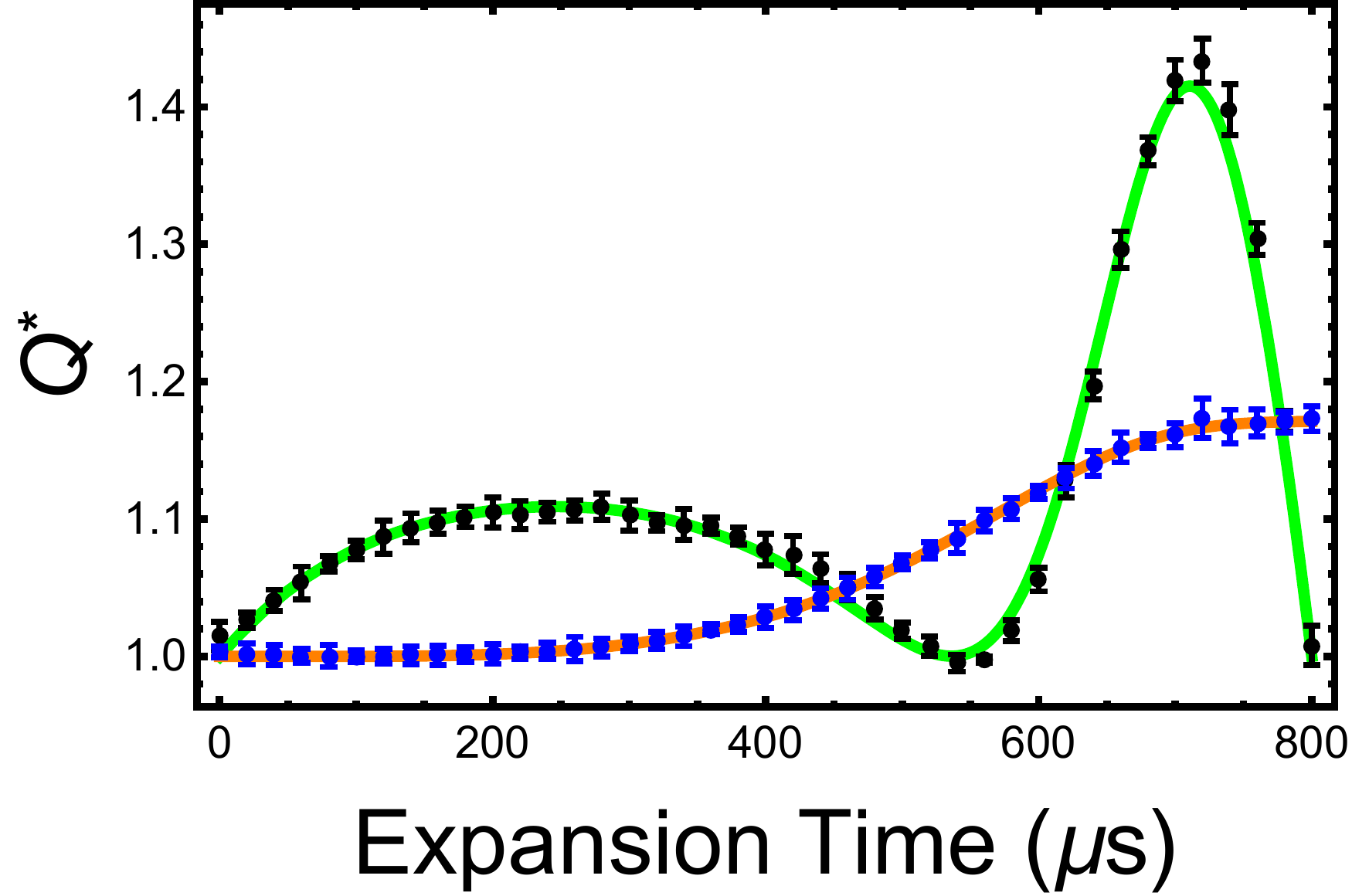}}
        {\includegraphics[width = 0.3 \textwidth]{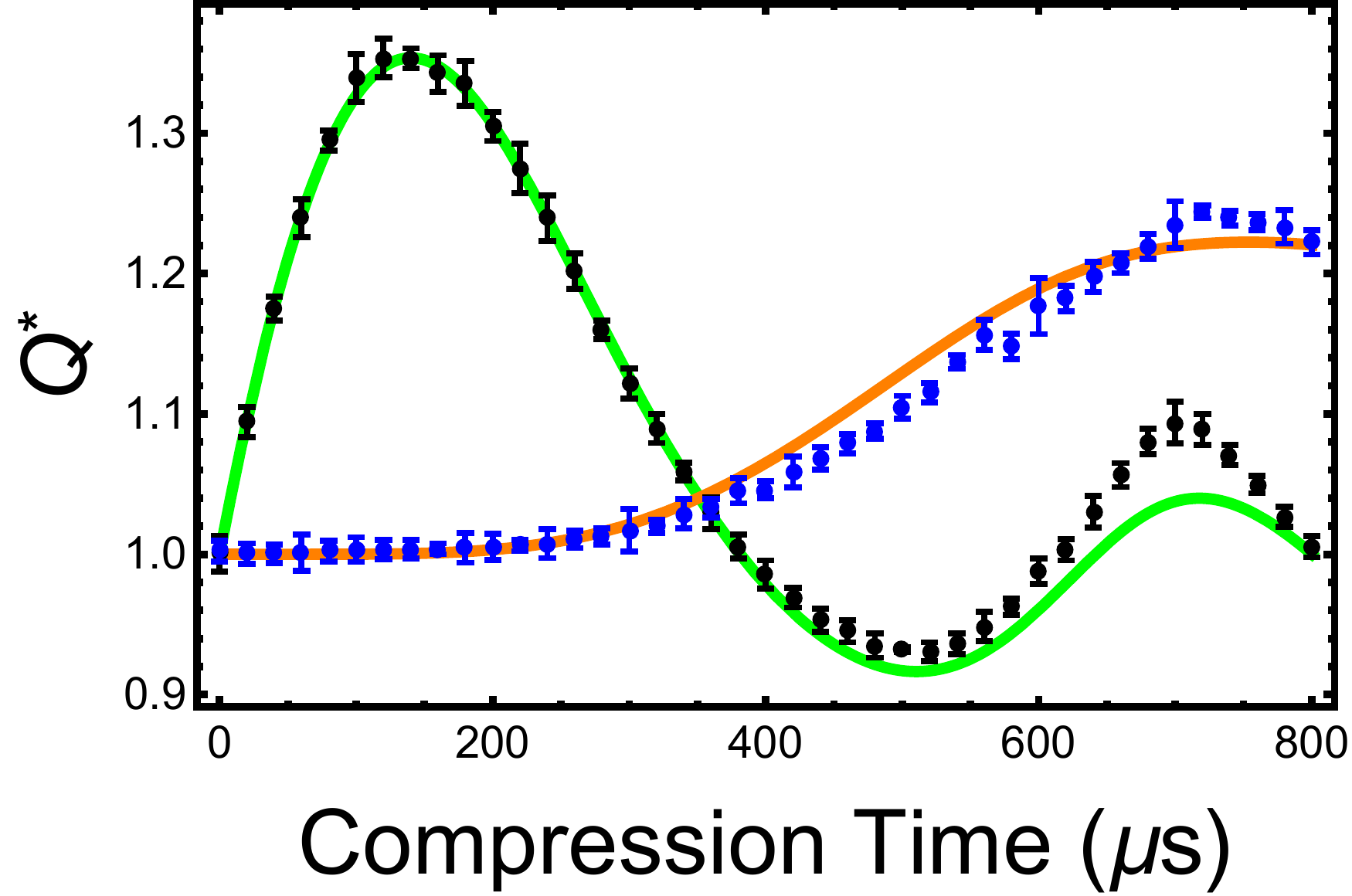}}
\caption{\label{ECNU}  {\bf Friction-free superadiabatic strokes.} Left: Experimental setup. The optical dipole trap is specially designed for a flexible control of the trap frequencies and consists of two orthogonal far-off resonance laser beams.  One beam is focused only in one spatial direction (x), i.e., providing confinement along this axis. A second, perpendicular laser beam confines the 3D unitary Fermi gas  along the  y and z axes.  The frequency in the x direction mostly depends on the power of the first beam. Similarly,  the second  beam determines the frequencies  in the y and z directions. The frequency aspect ratio of the trap can be simply controlled by precisely adjusting the power ratio of the two beams. Center: Evolution of the nonadiabatic factor $Q^*$ along an  expansion stroke with the unitary Fermi gas as a working substance. In the course of  the superadiabatic stroke,  $Q^*$ fluctuates reaching the value $Q^*=1$ at the end of the expansion. The green solid line corresponds  to the theoretical prediction and accurately matches the experimental data, in black. By contrast, a nonadiabatic expansion stroke $Q^*$ grows monotonically as a function of time and acquires  final values $Q^*>1$ associated with quantum friction. The theoretical prediction is plotted as a red solid line, in agreement with the experimental data, in blue. See \cite{Deng18} for details of the experiment and the parameters used.
Right:  Evolution of the nonadiabatic factor $Q^*$ along a superadiabatic compression stroke with the unitary Fermi gas as a working substance. The final value of $Q^*=1$ in a superadiabatic compression but takes values $Q^*>1$ in a general nonadiabatic stroke. Small discrepancies are observed in the transient fluctuations between the theoretical prediction (solid lines) and experimental data.
}
\end{center}
\end{figure}
Superadiabatic strokes  have been demonstrated in the laboratory using a unitary Fermi gas confined in an anisotropic harmonic trap as a working medium \cite{Deng18pra,Deng18}.
The Hamiltonian describing the system is 
\beqa
\hat{\mathcal{H}}(t)=\sum_{\sigma=\uparrow,\downarrow}\int d^3{\bf r}\hat{\psi}_\sigma^\dag({\bf r})\left[-\frac{\hbar^2}{2m}\nabla^2+V({\bf r},t)\right]\hat{\psi}_\sigma^\dag({\bf r})+g\hat{\psi}_\uparrow^\dag({\bf r})\hat{\psi}_\downarrow^\dag({\bf r})\hat{\psi}_\downarrow({\bf r})\hat{\psi}_\uparrow({\bf r})\ ,
\eeqa
where
\beqa
V({\bf r},t)=\frac{1}{2}m\left(\om_x^2x^2+\om_y^2y^2+\om_z^2z^2\right)\ .
\eeqa
Here, $\hat{\psi}_\sigma({\bf r})$ denotes the annihilation field operator for the spin state $\sigma=\uparrow,\downarrow$, $g$ is the coupling strength of the s-wave interactions and $\om_j$ ($j=x,y,z$) the trap frequencies along the different axis. 
For finite value of the interaction strength $g$, the system lacks scale invariance. However, the unitary limit can be reached by tuning the interaction strength, i.e., via a Feshbach resonance.
The system is then described by a nonrelativistic conformal field theory, with an associated scaling symmetry \cite{CD}, that can be exploited to implement superadiabatic  expansion and compression strokes, as those used in a quantum Otto cycle.  Their implementation in the laboratory was reported in \cite{Deng18}, that we briefly summarize next.  The spin-$1/2$ unitary Fermi gas was prepared using a balanced mixture of ${}^6$Li fermions in the lowest hyperfine states. By bringing the atomic cloud at resonance,  the scattering length governing the pairwise interactions can be tuned to exceed the interparticle distance and reach the unitary regime.  A sketch of the experimental setup is shown in Figure \ref{ECNU}. The effectively harmonic confinement in three dimensions was specially designed for an accurate and easy control of the trapping frequencies and their modulation in time. The theoretical analysis presented in the preceding sections for the isotropic trap can be extended to account for the anisotropy of the trap  \cite{Deng18}. Indeed, despite the anisotropy of the trap, it is possible to engineer an isotropic expansion factor $b(t)$ that is common to all axes (x, y and z). Figure \ref{ECNU}b-c report the time evolution of the nonadiabatic factor during an expansion and a compression, respectively. In both cases, the final value of the  nonadiabatic factor $Q^*$ reduces to the identity upon the completion of the superadiabatic stroke. By contrast, a generic nonadiabatic stroke results in values of $Q^*(\tau)>1$, that are associated with quantum friction.

Note that the above analysis does not make reference to the isochoric  strokes, that are assumed to be ideal (friction free) leading to the perfect thermalization of the working substance.  
By combining the superadiabatic expansion and compression demonstrated in \cite{Deng18} with the isochoric heating and cooling strokes, one can then envision the realization of a superadiabatic Otto cycle.

\section{Shortcuts to adiabaticity by counterdiabatic driving}
In the preceding sections we have focused on the use of quantum fluids  as a working medium.
The realization of STA in this context is greatly simplified due to the presence of scale invariance. In what follows we introduce  a universal technique to design STA in arbitrary quantum systems. It involves acting on a system Hamiltonian (e.g., describing the working medium) with auxiliary controls. Determining the later requires knowledge the spectral properties of the system, which maybe unavailable in complex many-body cases such as the quantum fluids we have discussed.

Consider a time-dependent Hamiltonian $\hat{H}(t)$ with instantaneous eigenvalues $\{\varepsilon_n(t)\}$ and eigenstates $\{|n(t)\ra\}$, so that
\beqa
\hat{H}(t)|n(t)\ra=\varepsilon_n(t)|n(t)\ra\ .
\eeqa
We pose the problem of driving an initial state  $|n(0)\ra$ to $|n(\tau)\ra$ in a given finite time, $\tau$, generally short enough for the dynamics to remain adiabatic. A technique which achieves this goal is the so-called counterdiabatic driving (CD) technique \cite{DR03,DR05}, also known as transitionless quantum driving \cite{Berry09}.

Whenever $\hat{H}(t)$ is slowly-varying,  the adiabatic approximation provides an approximate solution $|\psi_n^{\rm ad}(t)\ra$ of the time-dependent Schr\"odinger equation, i.e., 
\beqa
i\hbar\frac{d}{dt}|\psi_n^{\rm ad}(t)\ra\approx \hat{H}(t)|\psi_n^{\rm ad}(t)\ra\ .
\eeqa
Specifically, when the system is initialized in the the $n$-th eigenstate $|n(0)\ra$, the evolution follows the instantaneous eigenstate 
$|n(t)\ra$ as
\beqa
\label{adiabsol}
|\psi_n^{\rm ad}(t)\ra=\exp\left(-i\int_0^t\frac{\varepsilon_n(s)}{\hbar}ds-\int_0^t\la n(s)|\frac{d}{ds}|n(s)\ra ds \right)|n(t)\ra\ ,
\eeqa
where the exponential term includes the dynamical and  geometric phases  \cite{Berry84}.

The central goal  of CD is to find the so-called  counterdiabatic  Hamiltonian $\hat{H}_{\rm CD}$ such that  the adiabatic approximation $|\psi_n^{\rm ad}(t)\ra$ to the dynamic generated by $\hat{H}$ 
becomes the exact solution of the time-dependent Schr\"odinger equation  with $\hat{H}_{\rm CD}$
\beqa
i\hbar\frac{d}{dt}|\psi_n^{\rm ad}(t)\ra=\hat{H}_{\rm CD}|\psi_n^{\rm ad}(t)\ra\ ,
\eeqa
without the requirement of slow driving.

We assume that $\hat{H}_{\rm CD}$ is self-adjoint and look for the unitary  time-evolution operator, $\hat{U}_{\rm CD}(t,t'=0)$ that guides the dynamics through the  adiabatic reference trajectory
\beqa
|\psi_n^{\rm ad}(t)\ra=\hat{U}_{\rm CD}(t,0)|n(0)\ra
\eeqa
for all $|n(0)\ra$. As a result, $\hat{U}_{\rm CD}(t,0)$ also obeys the time-dependent Schr\"odinger equation
\beqa
i\hbar\frac{d}{dt}\hat{U}_{\rm CD}(t,0)=\hat{H}_{\rm CD}\hat{U}_{\rm CD}(t,0)\ .
\eeqa
The desired Hamiltonian can then be obtained from the time-evolution operator using the identity 
\beqa
\label{hcdu}
\hat{H}_{\rm CD}=i\hbar\left[\frac{d}{dt}\hat{U}_{\rm CD}(t,0)\right]\hat{U}_{\rm CD}(t,0)^\dag\  .
\eeqa
By construction, the time evolution operator is given by
\beqa
\hat{U}_{\rm CD}(t,0)=\sum_n |\psi_n^{\rm ad}(t)\ra\la n(0)|\ ,
\eeqa
whence it follows that the counterdiabatic Hamiltonian can be expressed as the sum 
\beqa
\hat{H}_{\rm CD}(t)&=&\hat{H}(t) + \hat{H}_1(t)
\eeqa
of the uncontrolled system Hamiltonian $\hat{H}(t)$ and an auxiliary counterdiabatic term
\beqa
\label{eqH1}
 \hat{H}_1(t)=i\hbar\sum_n\left(|d_tn\ra\la n|   -\la n|d_t n\ra |n\ra\la n|\right)\ .
\eeqa
The first term in the right hand side in Eq. (\ref{eqH1}) is responsible for suppressing excitations away from the $n$-th mode while the second one accounts for the Berry phase.
We notice that whenever the spectrum of the driven Hamiltonian $\hat{H}(t)$ is nondegenerate, we can rewrite the auxiliary term as
\beqa
\hat{H}_1=i\hbar\sum_n\sum_{m\neq n}\frac{|m\ra\la m|\dot{\hat{H}}|n\ra\la n|}{\varepsilon_n-\varepsilon_m}\ ,
\eeqa
where the sum is restricted to values of $m\neq n$ as the diagonal term is cancelled by the term that generates the Berry phase. This expression is physically very intuitive as it suggests that under the condition for adiabaticity the counterdiabatic driving explicitly vanishes, as it should. Further, it shows that the counterdiabatic term is off-diagonal in the energy eigenbasis of the system Hamiltonian $\hat{H}$. One can thus expect that the physical implementation of the the full auxiliary term might be challenging in the laboratory, as it requires carefully tuned matrix elements $|m\ra\la n|$.

The experimental demonstration of the CD technique has by now  been reported  in a range of platforms for quantum technologies in  quantum systems with a simple energy spectra \cite{expCD1,expCD2,expCD3,expCD4}. It can be extended to many-body spin systems undergoing a quantum phase transition, at the cost of implementing $n$-body interactions \cite{DRZ12}, which are necessary to suppress the universal formation of excitations and defects \cite{DZ14}.  

We close this section by emphasizing the nonadiabatic nature of shortcuts to adiabaticity, such as the driving protocols engineered using the counterdiabatic driving technique. We recall that $|\psi_n^{\rm ad}(t)\ra$ is the adiabatic approximation to the TDSE associated with $\hat{H}$. At the same time, it is the exact solution of the TDSE associated with $\hat{H}_{\rm CD}$. Given that $[\hat{H},\hat{H}_{\rm CD}]=[\hat{H},\hat{H}_{1}]\neq 0$, these two Hamiltonians do not share a common spectrum and cannot be diagonalized in the same basis. Said differently, $|\psi_n^{\rm ad}(t)\ra$ is diagonal in the eigenbasis of $\hat{H}$ but not in that of $\hat{H}_{\rm CD}$.
As a result, we conclude that $|\psi_n^{\rm ad}(t)\ra$ describes a nonadiabatic trajectory with respect to the instantaneous full counterdiabatic Hamiltonian, including transitions among its instantaneous eigenstates.

\section{Cost of counterdiabatic driving}

Provided an arbitrary counterdiabatic driving can be implemented, the duration of a STA can be made arbitrarily short. 
One can try to quantify the cost of implementing a CD scheme by analyzing the amplitude of the auxiliary driving fields as function of the (prescheduled) duration of the process $\tau$. 
Rescaling the time  of evolution $s=t/\tau\in[0,1]$, the spectral decomposition of the system Hamiltonian  can be written as $\hat{H}(s)=\sum_n\varepsilon_n(s)\hat{P}_n(s)$ in terms of the  projector 
$\hat{P}_n(s)=|n(s)\ra\la n(s)|$, where $|n(s)\ra$ is the instantaneous energy eigenstate.
The auxiliary counterdiabatic term can then be written as
\beqa
\hat{H}_1=\frac{i\hbar}{\tau}\sum_n \frac{d\hat{P}_n(s)}{ds}\hat{P}_n(s)\ .
\eeqa
Demirplak and Rice \cite{DR08} used  that the Hilbert-Schmidt norm  of the auxiliary term $\hat{H}_1$ to quantify the time-energy cost of counterdiabatic driving, finding
 \beqa
\|\hat{H}_1\|^2=\frac{\hbar^2}{2\tau^2}\sum_n\tr\dot{\hat{P}}_n^2 \ .
\eeqa
The authors also considered the time-integral of the norm, an analysis that has been further elaborated in \cite{Zheng16,CD17}.

An alternative characterization of the time-energy cost of STA resorts to the study of the energy fluctuations involved along the process \cite{DRZ12}.  In particular, the energy variance 
\beqa
\Delta H_{\rm CD}^2= \la \hat{H}_{\rm CD}^2\ra -\la \hat{H}_{\rm CD}\ra^2
\eeqa
is constrained by a time-energy uncertainty relation, and provides and upper bound to the speed of evolution in Hilbert space. 
For the total  Hamiltonian of the system given by the sum of the system Hamiltonian and the auxiliary control fields, it was found that the energy variance equals  the second moment of the control field. In particular, for the driving of a single eigenstate 
\beqa
\label{varH1}
\Delta H_{\rm CD}^2(t) = \sum_np_n^0\la n(t) |\hat{H}^2_1(t)| n(t)\ra\ .
\eeqa
In turn, the later acquires a  geometric interpretation in terms of the fidelity susceptibility $\chi_f^{(n)}(\lambda)$ 
that rules the decay of the overlap between an eigenstate of the Hamiltonian $\hat{H}(t)= \hat{H}[\lambda(t)]$ and the adiabatically continued eigenstate   under a small variation of the parameter $\lambda$ \cite{Gu10}
\beqa
|\la n( \lambda) | n(\lambda+\delta) \ra|^2 = 1- \delta^2 \chi_f^{(n)}(\lambda)+\mathcal{O}(\lambda^3)\ .
\eeqa
This is indeed the case as \cite{DRZ12}
\beqa
\la n(t) |\hat{H}^2_1(t)| n(t)\ra &=& \dot{\lambda}^2 \chi_f^{(n)}(\lambda)\nonumber\\
&=&\dot{\lambda}^2 \sum_{m\neq n} \frac{|\la m(\lambda) |\frac{d}{d\lambda} \hat{H}_0|n(\lambda)\ra|^2}{|\varepsilon_m - \varepsilon_n|^2}\ .
\eeqa

More recently, the thermodynamic cost of STA was analyzed by studying quantum work fluctuations \cite{Funo17,BT17}. For systems undergoing unitary dynamics, a possible definition of the work involves  in  driving the system from an initial Hamiltonian $\hat{H}(0)$ to a final one $\hat{H}(t)$ requires two energy measurements, one at the beginning of the process  ($t'=0$) and a second one upon its completion ($t'=t$)  \cite{Tasaki,Kurchan,fluctuation1}. 
Denoting the spectral decomposition of the instantaneous CD Hamiltonian by 
$\hat{H}_{\rm CD}(t)=\sum_n E_n(t) |E_n(t)\ra \la E_n(t)|$, the explicit expression for the work probability distribution $P[W(t)]$ associated with the CD is given by
\beqa
\label{P(W)}
P[W(t)]:=\sum_{k,n}p_n^0p_{n\rightarrow k}^t\delta[W(t)-(E_k(t)-E_n(0))]\ .
\eeqa
The probability for the initial state $\hat{\rho}$ to be found in the $n$-th eigenmode is thus given by $p_n^0=\la E_n(0)|\hat{\rho}|E_n(0)\ra$ while the transition probability from the $n$-th mode at $t=0$ to the $k$-th mode at time $t$ is given by $p_{n\rightarrow k}^t=|\la E_k(t)|\hat{U}_{\rm CD}(t,0)|E_n(0)\ra|^2$. 
In the implementation of CD, one is generally interested in the case in which the auxiliary control field $\hat{H}_1$ vanishes at the beginning and end of the driving protocol  so that $\hat{H}_{\rm CD}=\hat{H}$  at $t=\{0,\tau\}$. As a result, at these two instances of time (and only then), $E_n(t)=\varepsilon_n(t)$ and $|E_n(t)\ra=|\varepsilon_n(t)\ra$. 

We first note the work statistics in the adiabatic limit. Then, $H_1$ strictly vanishes, $\hat{H}_{\rm CD}(t)=\hat{H}(t)$, and  the transition probability becomes the Kronecker delta, $p_{n\rightarrow k}^t=\delta_{k,n}$  for all $t$. 
The work probability distribution under adiabatic evolution is thus
\beqa
P_{\rm ad}[W(t)]
=\sum_{n}p_n^0\delta[W(t)-W_{\rm ad}^{(n)}(t)]\ ,
\eeqa
where $W_{\rm ad}^{(n)}(t):=\epsilon_{n}(t)-\epsilon_{n}(0)$ is the work cost along the adiabatic trajectory of the $n$-th eigenmode. In particular, the mean work is given by
\beq
\la W(t)\ra_{\rm ad}:=\int \mathrm{d}W P_{\rm ad}[W] W(t)=\sum_{n}p_n^0 [\varepsilon_n(t)-\varepsilon_n(0)]\ .
\eeq

In a STA, given that upon completion of the protocol $\hat{H}(\tau)$ remains constant and $\hat{H}_1(\tau)$ vanishes, the work probability distribution reads
\beqa
P_{\rm CD}[W(\tau)]=P_{\rm ad}[W(\tau)]\ .
\eeqa
Therefore, CD successfully reproduces the work statistics under slow driving, and in this sense, it has no thermodynamic cost.

Along the STA ($0<t<\tau$), however, CD does modify the work statistics, i.e., $P_{\rm CD}(W)$ differs from $P_{\rm ad}[W(t)]$. Nonetheless, it satisfies two remarkable properties.
First, the mean work identically matches the adiabatic value 
\beqa
\label{equality1}
\la W(t)\ra=\la W(t)\ra_{\rm ad}\ .
\eeqa
Said differently, the mean work done by the auxiliary counterdiabatic term vanishes for all $0\leq t\leq\tau$.

Second, CD enhances work fluctuations. If the system Hamiltonian $\hat{H}(\lambda)$  depends explicitly on a set of parameters $\lambda=(\lambda^1,\dots,\lambda^N)$, the excess of the work variance over the adiabatic value is precisely given by
\beqa
\mathrm{Var}[W(t)]-\mathrm{Var}[W(t)]_{\rm ad}=\hbar^2  \sum_np_n^0g_{\mu\nu}^{(n)}\dot{\lambda}^\mu\dot{\lambda}^\nu\ .
\label{wvarq}
\eeqa
The term in the right hand side is the  average of the quantity $g_{\mu\nu}^{(n)}$
weighted with the occupation $p_n^0$ of the energy levels following the first projective energy measurement.
As $g_{\mu\nu}^{(n)}$ is the real part of the quantum geometric tensor $Q_{\mu\nu}^{(n)}$ of the $|n(t)\ra$-state manifold  \cite{PV80},
\beqa
\label{eqqgt}
Q_{\mu\nu}^{(n)}:=\la \partial_{\mu}n(t)|[1-|n(t)\ra\la n(t)|]|\partial_{\nu}n(t)\ra\ , 
\eeqa
the broadening of the work distribution is dictated by the geometry of the Hilbert space. When the system Hamiltonian is modulated by a single parameter, $g_{\mu\mu}^{(n)}=\chi_f^{(n)}$.
Using the identity (\ref{eqqgt}) and analyzing the time-average excess of work fluctuations, it is also possible to derive work-time uncertainty relations \cite{Funo17,BT17}.  The excess of work fluctuations induced by CD, Eq. (\ref{wvarq}), has  recently been verified experimentally using a superconducting Xmon qubit \cite{Hangzhou}.

We conclude this section by pointing out that  a common feature among the different approaches to quantify the cost of STA is that $\|\hat{H}_1\|^2$, $\Delta H_{\rm CD}^2(t)$ and $\mathrm{Var}[W(t)]-\mathrm{Var}[W(t)]_{\rm ad}$ all diverge as $1/\tau^2$ as the protocol duration $\tau$ is reduced.  This is consistent with the fact that the above quantities are ultimately quadratic in the auxiliary control term $\hat{H}_1$. This scaling with the  duration $\tau$ is specific of counterdiabatic driving and can differ from that in related STA protocols in which the evolution is generated by a unitarily equivalent Hamiltonian, an approach often referred to as local counterdiabatic driving \cite{expCD4,Deng18}, see as well \cite{Zheng16}. In addition, counterdiabatic driving generates coherence in the instantaneous energy eigenbasis along the STA. In the presence of decoherence, the work input and output associated with the compression and expansion strokes can be thus affected \cite{Levy18}.

\section{Choice of working medium, many-particle quantum effects and quantum supremacy}

We have seen that by using STA, an Otto cycle can be operated in finite-time without friction (assuming friction associated with the heating and cooling strokes to be negligible). In this section we focus on the finite-time thermodynamics in the presence of friction, e.g., with an efficiency below the maximum value. The  performance of the engine can then exhibit a dependence on the nature of the working medium. Which kind of substance would be optimal? 

Jaramillo et al. \cite{Jaramillo16} considered a fixed number of particles $\N$ for the working medium and compared the performance of one single many-particle heat engine with an ensemble of $\N$ single-particle heat engines. In doing so, it was shown that it is possible to find scenarios characterized by quantum supremacy, a many-particle quantum enhancement of the performance  with no classical counterpart.

This effect was illustrated with a working medium consisting of $\N$  bosons  in an effectively one-dimensional harmonic trap and subject to inverse-square pairwise interactions \cite{Calogero71,Sutherland71},
\begin{eqnarray}
 \hat{H}(t)=\sum_{i=1}^{\N}\left[-\frac{\hbar^2}{2m}\frac{\partial^2}{\partial z_i^2}+\frac{1}{2}m\omega(t)^2 z_{i}^{2}\right]+\frac{\hbar^2}{m}\sum_{i<j}\frac{\lambda(\lambda-1)}{(z_i-z_j)^2}\ ,
 \label{eq:csm}
\end{eqnarray}
where  $\lambda\geq 0$ is the  interaction strength. 
This instance of (\ref{Hsu11}) is the (rational) Calogero-Sutherland model that reduces to an ideal Bose gas for $\lambda=0$ and to the Tonks-Girardeau gas (hard-core bosons) for $\lambda=1$ \cite{Girardeau60,GWT01},  the thermodynamics of which   is equivalent to that of polarized fermions. For arbitrary $\lambda$, the Calogero-Sutherland can be interpreted as an ideal gas of particles obeying generalized-exclusion statistics \cite{Haldane91,Wu94,MS94}.

The  comparative performance in both scenarios can be assessed via the power and efficiency ratios, defined respectively by
\beqa
\frac{P^{(\N,\lambda)}}{\N\times P^{(1,\lambda)}}\ , \quad  \frac{\eta^{(\N,\lambda)}}{\eta^{(1,\lambda)}}\ .
\eeqa

In an ample regime of parameters, it was shown that both ratios can simultaneously increase and surpass unity. The enhancement of the efficiency is maximum
for an ideal Bose gas ($\lambda=0$). While the effect is robust in the presence of interactions, the ensemble of single-particle engines can however outperform the many-particle one for moderate values of $\lambda$, e.g., for hard-core bosons (fermions). 

Bengtsson et al. have identified a similar many-particle boost  in the average work output  in the adiabatic limit of a quantum Szilard engine, when an attractive Bose gas is chosen as a working medium \cite{Bengtsson18}. 
 Zheng and Poletti have uncovered quantum statistical effects in the performance of an adiabatic Otto cycle for a given working medium (ideal Bose and Fermi gases), confined in non-harmonic traps \cite{ZP15}.  It has further been shown that the performance of an Otto cycle can also be enhanced whenever the working substance exhibits critical behavior \cite{Campisi16}.  An analysis  of this cycle with a working medium exhibiting many-body localization has also been reported \cite{Nicole17}. 
 
Overall, the optimal choice of the working substance for a given quantum cycle seems to be largely unexplored. Further, this choice may depend on the assessment of the performance as the characterization of a quantum thermodynamic cycle does not necessarily carry over many cycles when work is measured via projective energy measurements \cite{Watanabe17}. 
Nonetheless, it seems that a rich variety of scenarios can be found in which many-particle quantum effects can lead to an improvement of the performance over classical many-particle cycles as well as ensembles of single-particle quantum machines.

\section{Quo vadis?}

The time of writing is  particularly exciting  concerning experimental progress towards the realization of quantum heat engines at the nanoscale.  Using trapped-ion technology a single atom heat engine has been realized \cite{Rossnagel15}.  A  quantum heat engine based on an Otto cycle has been implemented using a spin-1/2 system and nuclear magnetic resonance techniques \cite{Peterson18}. 
Quantum refrigerators have as well been reported using superconducting circuits \cite{Yen16} and trapped ions \cite{Maslennikov17}.
The possibility of combining these setups with superadiabatic strokes such as those reported in \cite{Deng18} suggest that the realization of friction-free thermal machines may be feasible in the near future.
This prospect is enhanced by the fact that  superadiabatic machines are not necessarily restricted to the quantum domain and can be envisioned in classical systems as well, such as colloidal heat engines \cite{Martinez17}.

The current description of quantum heat engines based on a thermodynamic  cycle consisting of multiple strokes disregards the explicit mechanism for work outcoupling, this is, the coupling between the working medium and an external work storage. As a result, work is merely associated with a change of a  parameter of the working-medium Hamiltonian, such as the frequency of the trap in the quantum Otto cycle. A direct  implementation of such scheme can be envisioned  via  external active control, inducing a given modulation in time  of the  Hamiltonian parameter.

Useful heat engines are expected to run autonomously without the requirement for external controls. With an eye on experimental implementations, it is therefore necessary to transcend the current description. This is likely to require a quantum description of the full engine including the fuel, working substance, and work storage. Such description will no doubt be less universal, but is likely to be a highly fruitful pursuit.  Encouraging steps in this direction include an account of the dynamics of quantum machine that lifts a weight against gravitational field \cite{Gilz16} and tilted potentials \cite{Teo17}, an autonomous absorption refrigerator in atom-cavity systems \cite{Mitchison16} as well as an autonomous rotor heat engine \cite{Roulet17}, discussed in \chqre. 

We anticipate that the engineering of superadiabatic quantum machines will be at reach in the presence of external active control. By contrast, their autonomous counterpart is likely to prove more challenging. The ultimate cost of STA may therefore arise form its practical implementation. The study of STA has however shown that they can be implemented following simple principles, such as slow driving at the beginning and end of a process \cite{Chen10,delcampo13}.

In addition, our discussion of superadiabatic heat engines has focused on the Otto cycle. This choice is motivated by the presence of strokes governed by unitary dynamics, for which STA techniques are well developed.  In general thermodynamic cycles (e.g., the Carnot cycle) the dynamics is open (in contact with a heat reservoir) in all stages. The engineering of superadiabatic engines based on such cycles would require the use of STA for open quantum systems, which is an open problem. This challenge would also be faced in continuously driven cycles.

\bigskip

\section{Acknowledgements}
This chapter reports on joint work with Shuoming An, Mathieu Beau, Cyril Chatou,  Ivan Coulamy,   Pengpeng Diao, Ken Funo, John Goold, Juan Diego Jaramillo, Kihwan Kim, Fang Li, Mauro Paternostro, Marek M. Rams, Masahito Ueda,  Shi Yu and Jing-Ning Zhang, and Wojciech H. Zurek.  It has further benefited from discussions with Obinna Abah, Sebastian Deffner, Luis Pedro Garc\'ia-Pintos,  Fernando J. G\'omez-Ruiz, Jiangbin Gong, Christopher Jarzynski, Ronnie Kosloff, Peter H\"anggi,  Eric Lutz, Doha M. Mesnaoui, Victor Mukherjee,  Jos\'e Pascual Palao, Dario Poletti, Stuart A. Rice, Peter Talkner, B. Prasanna Venkatesh,  Gentaro Watanabe, and Zhenyu Xu. We acknowledge funding support from the John Templeton Foundation, and UMass Boston (project P20150000029279).

\end{document}